\documentclass{aa}
\usepackage{graphicx}
\usepackage{txfonts}

\begin{document}
  
  \title{The structure of the protoplanetary disk surrounding three young
    intermediate mass stars \thanks{Based on observations collected at
      the European Southern Observatory, Paranal, Chile (Proposal ID:
      077.C-0521, 077.C-0263)} }
  \subtitle{II. Spatially resolved dust and gas distribution}
  
  \author{D. Fedele \inst{1,2,3}, 
    M. E. van den Ancker \inst{3}, 
    B. Acke \inst{4}, 
    G. van der Plas \inst{3,5},
    R. van Boekel \inst{1},
    M. Wittkowski \inst{3}, 
    Th. Henning \inst{1},
    J. Bouwman \inst{1},
    G. Meeus \inst{6},
    and
    P. Rafanelli \inst{2}
}  
\offprints{D. Fedele}
  \institute{Max Planck Institut f\"ur Astronomie, K\"onigstuhl 17, D-69117 Heidelberg, Germany\\
    \email{dfedele@mpia.de}
    \and
    Dipartimento di Astronomia, Universit\'a degli studi di Padova, Vicolo dell'Osservatorio 2, 35122 Padova, Italy
    \and
    European Southern Observatory, Karl Schwarzschild Strasse 2, D-85748 Garching bei M\"unchen, Germany
    \and 
    Instituut voor Sterrenkunde, KU Leuven, Celestijnenlaan 200B, 3001 Leuven, Belgium
    \and
    Sterrenkundig Instituut 'Anton Pannekoek', University of Amsterdam, Kruislaan 403, 1098 SJ Amsterdam, The Netherlands
    \and
    Astrophysikalisches Institut Potsdam, D-14482 Potsdam, Germany
  }
  
  \date{}
  \titlerunning{The structure of protoplanetary disks}
  \authorrunning{Fedele et al.}
  
  \abstract
  {}
  {We present the first direct comparison of the distribution of the gas, as
    traced by the [\ion{O}{i}] 6300\AA\ emission, and the dust, as traced by
    the 10 $\mu$m emission, in the planet-forming region of proto-planetary
    disks around three intermediate-mass stars: HD 101412, HD 135344 B and HD
    179218.}
  {$N$-band visibilities were obtained with VLTI/MIDI. Simple geometrical
    models are used to compare the dust emission to high-resolution optical
    spectra in the 6300\,\AA\ [\ion{O}{i}] line of the same targets.}  
  {HD 101412 and HD 135344 B show compact ($<$ 2 AU) 10 $\mu$m emission while
    the [\ion{O}{i}] brightness profile shows a double peaked structure. The
    inner peak is strongest and is consistent with the location of the dust,
    the outer peak is fainter and is located at 5-10 AU. In both systems,
    spatially extended PAH emission is found. HD 179218 shows a double
    ring-like 10 $\mu$m emission with the first ring peaking at $\sim$ 1 AU
    and the second at $\sim$ 20 AU. The [\ion{O}{i}] emitting region is more
    compact, peaking between 3 -- 6 AU.}
  {The disks around HD 101412 and HD 135344 B appear strongly flared in the
    gas, but self-shadowed in the dust beyond $\sim$ 2 AU. The difference in
    the gas and dust vertical structure beyond 2 AU might be the first
    observational evidence of gas-dust decoupling in protoplanetary disks. The
    disk around HD 179218 is flared in the dust. The 10 $\mu$m emission
    emerges from the inner rim and from the flared surface of the disk at
    larger radii. No dust emission is detected between $\sim$ 3 -- 15 AU. The
    oxygen emission seems also to come from a flared structure, however, the
    bulk of this emission is produced between $\sim$ 1 -- 10 AU. This could
    indicate a lack of gas in the outer disk or could be due to chemical
    effects which reduce the abundance of OH -- the parent molecule of the
    observed [\ion{O}{i}] emission -- further away from the star. It may also
    be a contrast effect if the [\ion{O}{i}] emission is much stronger in the
    inner disk. We suggest that the three systems, HD 179218, HD 135344 B and
    HD 101412, may form an evolutionary sequence: the disk initially flared
    becomes flat under the combined action of gas-dust decoupling, grain
    growth and dust settling.}
  \keywords{Stars: pre-main sequence; circumstellar matter: protoplanetary
    disks}
  \maketitle

  \section{Introduction}
  Many pre-main-sequence stars are characterized by excess infrared emission
  above the stellar photospheric level which, depending on the evolutionary
  state of the system, may start in the near infrared (1 -- 5~$\mu$m) or at
  longer wavelengths. The dust distributed around the young star is
  responsible for this emission. The dust particles absorb a large fraction of
  the short wavelength stellar photons and re-emit them at infrared
  wavelengths. This dust is believed to be confined to a disk-like structure
  which forms in the early phase of star formation as a result of the
  conservation of the angular momentum of the parental cloud. Disk formation
  is followed by a longer phase of disk accretion during which disk material
  accretes onto the young star at a typical accretion rate of $10^{-7} -
  10^{-10}$ M$_{\sun}$ yr$^{-1}$. The interstellar dust and gas that forms the
  disk undergoes changes in its composition and size. Infrared surveys show
  that after a mean age of 3 Myr the inner part of the disk is cleared of
  dust. Viscous accretion, photo-evaporation and planet formation are the
  likely mechanisms responsible for this phenomenon (see review in Henning
  \cite{henning}).
  
  \noindent
  
  \begin{table*}
    \caption{Properties of the programme stars. The column "Meeus group" gives the
      classification in {\it flared} (group I) and {\it self-shadowed} (group
      II) disks by Meeus et al. (\cite{meeus}). Column ``[\ion{O}{i}] extent''
      reports the extent (minimum and maximum radius) of the [\ion{O}{i}]
      emitting region as derived in paper I. The disk position angle (PA) and
      inclination are taken from paper I.}
    \label{tab:prop}
    \centering
    \begin{tabular}{lllllllllll}
      \hline\hline
      Star        & RA      & DEC     & Sp.T & M$_{star}$   & F$_{12\mu m}$ & Distance & Meeus Group & PA & Inclination & [\ion{O}{i}] extent\\
                  & (J2000) & (J2000) &      & [M$_{\sun}$] & [Jy]        & [pc]     &             & [\degr] & [\degr]          & [AU]      \\     
      \hline
      HD 101412   & 11:39:44.46 & -60:10:27.7 & A0IIIe & 2.3 & 3.22       & 160      & II        &  --     & 30               & 0.15 - 10 \\ 
      HD 135344 B & 15:15:48.44 & -37:09:16.0 & F4Ve   & 1.7 & 1.59       & 140      & I         & 100     & 45               & 0.1 - 100 \\
      HD 179218   & 19:11:11.25 & +15:47:15.6 & B9e    & 2.7 & 23.4       & 240      & I         & 10      & 40               & 0.4 - 50  \\
      \hline\hline
    \end{tabular}
  \end{table*}

  A circumstellar disk is believed to be the locus where planet formation
  takes place. The large number of recently discovered exo-planetary systems
  and the variety of such systems raised many new questions about the
  structure and evolution of {\it protoplanetary} disks. Of particular
  interest for the understanding of disk evolution and planet formation is the
  coupling of gas and dust in disks. A main assumption underlying essentially
  all proto-planetary disk models is that dust and gas are thermally
  coupled. It is an open question whether this assumption holds on the disk
  surface and efforts have been made to improve disk models by taking into
  account the dust-gas decoupling (e.g. Kamp \& Dullemond \cite{kd}). While
  gas and dust are thermally coupled in the disk interior, in the low density
  environments of the disk surface layer, the two components may decouple.
 
  \smallskip
  \noindent

  The detailed structure of the disk surface temperature in the presence of
  gas-dust decoupling was studied by Jonkheid et al. (\cite{jonkheid}), Kamp
  \& Dullemond (\cite{kd}) and Nomura \& Millar (\cite{nomura}). Different
  heating/cooling processes act at different heights in the disk. In
  particular, very high in the atmosphere, with particle densities as low as
  {\it n} $< 10^5 {\rm cm}^{-3}$ (A$_V$ $\lesssim$ 10$^{-3}$ mag), the gas
  temperature is set by the balance of photoelectric heating and fine
  structure line cooling of neutral oxygen (Jonkheid et al. \cite{jonkheid},
  Kamp \& Dullemond \cite{kd}). In the upper layers of protoplanetary disks
  the gas temperature may exceed the dust temperature. At small radii ($<$ 50
  AU) the temperature of the gas above the disk photosphere may reach $\sim
  10^4$ K. At larger radii ($>$ 50 AU) the gas can become as hot as a few
  hundred Kelvin (Kamp \& Dullemond \cite{kd}). However, it is not obvious
  that such a hot and tenuous disk atmosphere can remain as a stable structure
  of a disk. The difference in gas and dust temperature as well as other
  processes (e.g. disk wind, disk evaporation, dust coagulation and settling)
  may lead to a physical decoupling of gas and dust in disks.

  \smallskip
  \noindent

  In this paper we present the first direct comparison of the dust and gas
  emission of three pre-main-sequence stars: \object{HD 101412}, \object{HD
    135344 B} and \object{HD 179218}. These are intermediate-mass (1.7 - 2.7
  M$_{\sun}$) stars belonging to both group I (flaring disk: HD 135344 B and
  HD 179218) and group II (flattened disk: HD 101412) according to the
  classification of Meeus et al. (\cite{meeus}). In a previous paper (van der
  Plas et al. \cite{plas}, hereafter paper I) we presented high resolution
  spectroscopy ($\frac{\lambda}{\Delta\lambda} = 77000$) of the optical
  [\ion{O}{i}] 6300.304\AA\ line with VLT/UVES of these three stars. Here we
  present $N$-band interferometric observations obtained with VLTI/MIDI aimed
  at spatially resolving the mid-infrared emitting region of the disk. We will
  also compare the thermal coupling between dust and gas in the disks
  surroundings our three targets by comparing the MIDI observations to a
  simple phenomenological model derived from the [\ion{O}{i}] data. The
  properties of the programme stars are reported in Table \ref{tab:prop}. The
  paper is organized as follow: section 2 briefly summarizes the results of
  paper I; section 3 contains information about observations and data
  reduction; in section 4 and 5 we report respectively an analysis of the
  interferometric measurements and the comparison with the optical data of
  paper I. The discussion and conclusion are given in section 6.
  
  \section{[\ion{O}{i}] emission: paper I}
  The [\ion{O}{i}] 6300\AA\ emission in young intermediate-mass stars is
  thought to be caused by photodissociation of the OH molecules present in the
  protoplanetary disk by stellar UV photons. A fraction of the resulting
  excited oxygen atoms is in the upper state ($^{1}D_2$) of the
  6300\AA\ transition. The [\ion{O}{i}] 6300\AA\ line is thus non-thermal and
  strongly sensitive to the UV radiation from the central star (Acke et
  al. \cite{acke05}). For this reason, the oxygen line traces only the surface
  layers of the disk that are directly exposed to the stellar radiation
  field.
  
  The Doppler broadened oxygen emission profile was translated into an amount
  of emission as a function of distance from the central star assuming
  Keplerian rotation of the gas. The resulting radial profile of the
  [\ion{O}{i}] emission is in agreement with the expected disk shapes as
  derived from their spectral energy distribution (SED) according to the
  phenomenological classification of Meeus et al. (\cite{meeus}). For all
  targets the oxygen emission starts at velocities corresponding to their dust
  sublimation radius and extends up to radii of 10 -- 90 AU. The [\ion{O}{i}]
  radial profile shows a double peak structure in the case of HD 101412 and HD
  135344 B, with a first stronger peak at $\sim$ 1 AU, and a second weaker
  peak at $\sim$ 5-10 AU. In HD 179218 the [\ion{O}{i}] has a single peak at
  $\sim$ 3 -- 6 AU. The normalized intensity [\ion{O}{i}] radial profiles for
  the three stars are shown in Figs. \ref{fig:OI_1}, \ref{fig:OI_2} and
  \ref{fig:OI_3}.

 \section{Observations and data reduction}  
  MIDI\footnote{www.eso.org/instruments/midi} (Leinert et al. \cite{leinert})
  is the mid-infrared (8 -- 13~$\mu$m) beam combiner of the VLT interferometer
  on Cerro Paranal, Chile. MIDI can combine the light of any pair of the 4 ESO
  VLT telescopes (UT, 8.2-m). MIDI allows spectrally resolved observations,
  that is, the signal is dispersed by either a prism ($\Delta \lambda /
  \lambda \approx 30$) or a grism ($\Delta \lambda / \lambda \approx
  300$). For our project we used the low resolution (prism) mode and two
  nearly perpendicular baselines: UT1--UT2 ({\it B} = 57m, PA = 26\degr) and
  UT3--UT4 ({\it B} = 62m, PA = 111\degr). Taking advantage of the earth
  rotation (which modifies the projection of the baseline on the sky) we
  observed our targets at different sidereal times, to optimize the coverage of
  the {\it(u,v)} plane. 

  \begin{figure*}[!ht]
    \centering
    \includegraphics[height=\textwidth, angle=90]{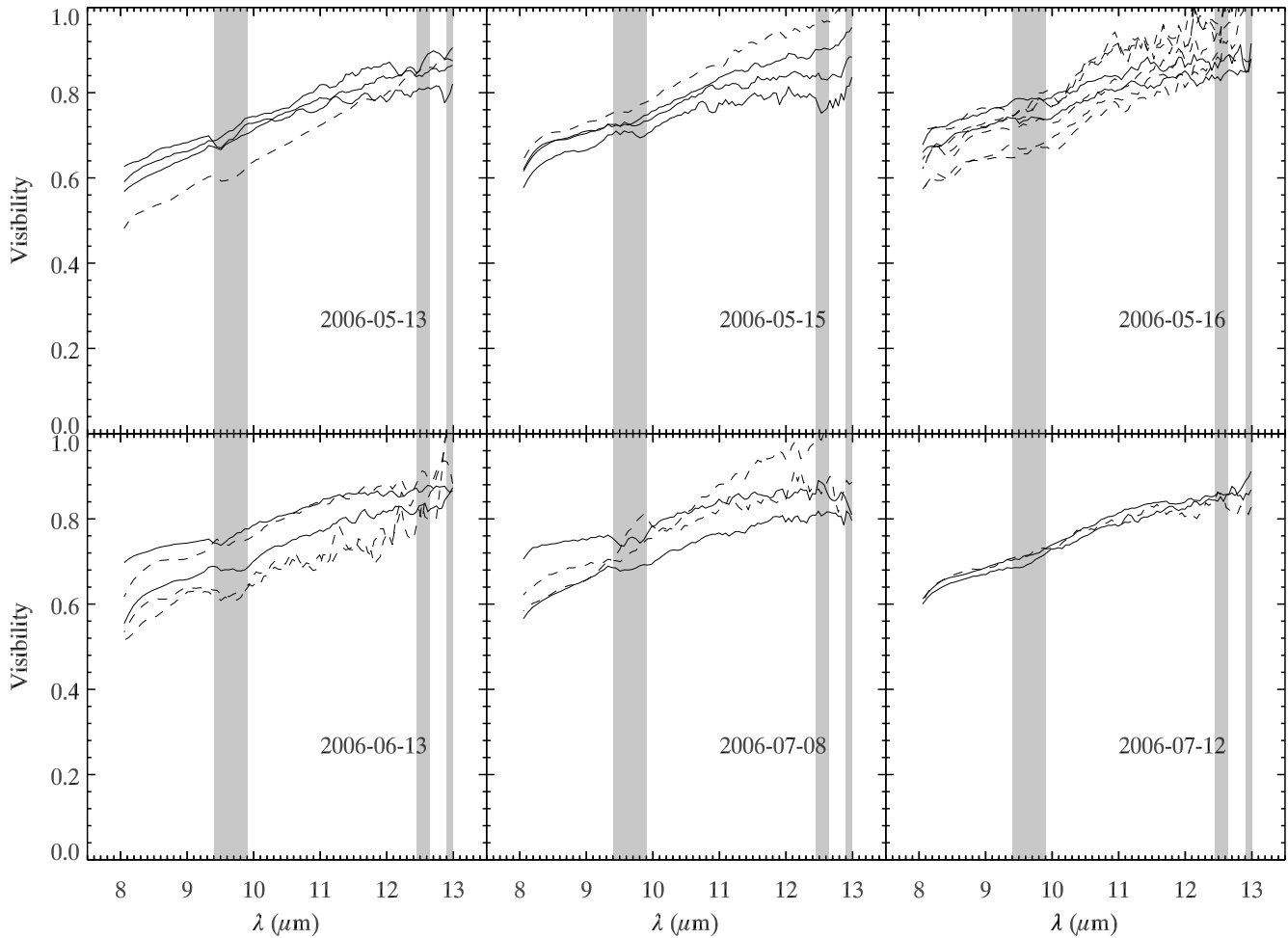}
    \caption{Transfer function of VLTI/MIDI. For each night, the different
      lines correspond to different calibrators observed with the prism in
      both modes: HS (solid line) and SP (dashed line). The overall shape of
      the transfer function, for a given mode, does not drastically change
      during the night. The vertical grey regions are characterized by strong
      atmospheric absorption bands. Residuals of the data reduction may be
      present within these regions.}
    \label{fig:tf}
  \end{figure*}
  \begin{table*}
    \caption{Summary of VLTI/MIDI observations. Column "Mode" indicates the
      beam combination mode of the interferometric signal: High-Sens (HS) or
      Sci-Phot (SP). Columns "$B_{\perp}$" and "PA" list the length of the
      projected baseline and its position angle (east of north).}
    \label{tab:log}
    \centering
    \begin{tabular}{lllllllll}
      \hline\hline
      Night & UT      & Mode & $\lambda/\Delta\lambda$ & Baseline& $B_{\perp}$
      & PA       & Airmass & Calibrator \\
      & [hh:mm] &      &                         &         & [m]   &
      [$\degr$]&         &            \\
      \hline
      \multicolumn{9}{c}{HD 101412} \\
      \hline  
      2006-05-15 & 23:38 - 00:02 & SP   & 30  & UT3-UT4 & 59  & 102 & 1.2 & HD 107446 \\
      2006-05-16 & 01:55 - 02:18 & SP   & 30  & UT3-UT4 & 62  & 110 & 1.3 & HD 107446 \\
      2006-06-13 & 00:18 - 00:34 & SP   & 30  & UT1-UT2 & 43  & 30  & 1.3 & HD 107446 \\
      \hline                                           
      \multicolumn{9}{c}{HD 135344 B} \\              
      \hline                                           
      2006-05-13 & 04:32 - 05:15 & HS   & 30  & UT3-UT4 & 62  & 112 & 1.0 & HD 129456 \\    
      2006-06-13 & 23:26 - 23:52 & HS   & 30  & UT1-UT2 & 56  & 27  & 1.3 & HD 129456 \\
      2006-07-12 & 03:14 - 03:47 & HS   & 30  & UT1-UT2 & 47  & 10  & 1.3 & HD 129456 \\
      \hline                                           
      \multicolumn{9}{c}{HD 179218} \\                
      \hline                                           
      2006-05-16 & 05:25 - 05:44 & SP   & 30  & UT3-UT4 & 58  & 128 & 1.7 & HD 188512 \\
      2006-06-13 & 06:42 - 06:56 & SP   & 30  & UT1-UT2 & 48  & 42  & 1.3 & HD 188512 \\
      2006-06-13 & 08:07 - 08:25 & SP   & 30  & UT1-UT2 & 53  & 40  & 1.5 & HD 188512 \\
      2006-07-08 & 06:45 - 07:03 & SP   & 30  & UT3-UT4 & 44  & 84  & 1.6 & HD 188512 \\
      \hline\hline
    \end{tabular}
  \end{table*}
  
  Two modes are available for the acquisition of the interferometric signal:
  ``High Sens'' and ``Sci Phot''. With the first mode, the photometric signal
  (i.e. the pure, not combined, 8 -- 13~$\mu$m spectrum of the source coming
  from the two telescopes) is acquired soon after the interferometric
  signal. In Sci Phot mode, the two signals are acquired at the same time. We
  used both modes depending on the N-band luminosity of the source. The log of
  MIDI observations is reported in Table \ref{tab:log}. The data were reduced
  with the standard data reduction software MIA+EWS
  v1.5.2\footnote{www.mpia-hd.mpg.de/MIDISOFT}\fnmsep\footnote{www.strw.leidenuniv.nl/~koehler/MIA+EWS-Manual}
  using the Expert Work Station (EWS). Detailed documentation on the reduction
  procedure can be found in Ratzka (\cite{ratzka}) and Jaffe (\cite {jaffe})
  or in the cited web pages.
      
  \smallskip
  \noindent

  Standard stars (\object{HD 107446}, \object{HD 129456} and \object{HD
    188512}) were observed close in time and in airmass to the science
  targets. The angular diameter of these stars is well known and they can be
  used as calibrators for the transfer function of the instrument. Other
  calibrators observed during these nights were used to check the stability of
  the transfer function. The error on the calibrated visibility of the science
  targets is the standard deviation resulting from calibration with different
  calibrator stars for each night. Fig. \ref{fig:tf} shows the transfer
  function (i.e. the instrumental visibility of the calibrators) for the six
  nights of observations. For each night, all the calibrators observed with
  the prism in both modes (HS and SP) are plotted. The overall shape of the
  transfer function with wavelength varies by less than 10\% during the
  night. The standard stars were also used to flux calibrate the mid-infrared
  spectrum of the science targets. Conversion factors from counts to Jy were
  computed from theoretical spectral energy distributions (SED). Energy
  distributions were derived by matching the spectral type of the calibrators
  to stars in the Cohen list (Cohen et al. \cite{cohen}) and scaling the
  spectrum with the ratio of their IRAS 12~$\mu$m fluxes. 
  
  \smallskip
  \noindent

\begin{figure}[!ht]
   \centering
   \includegraphics[width=9cm]{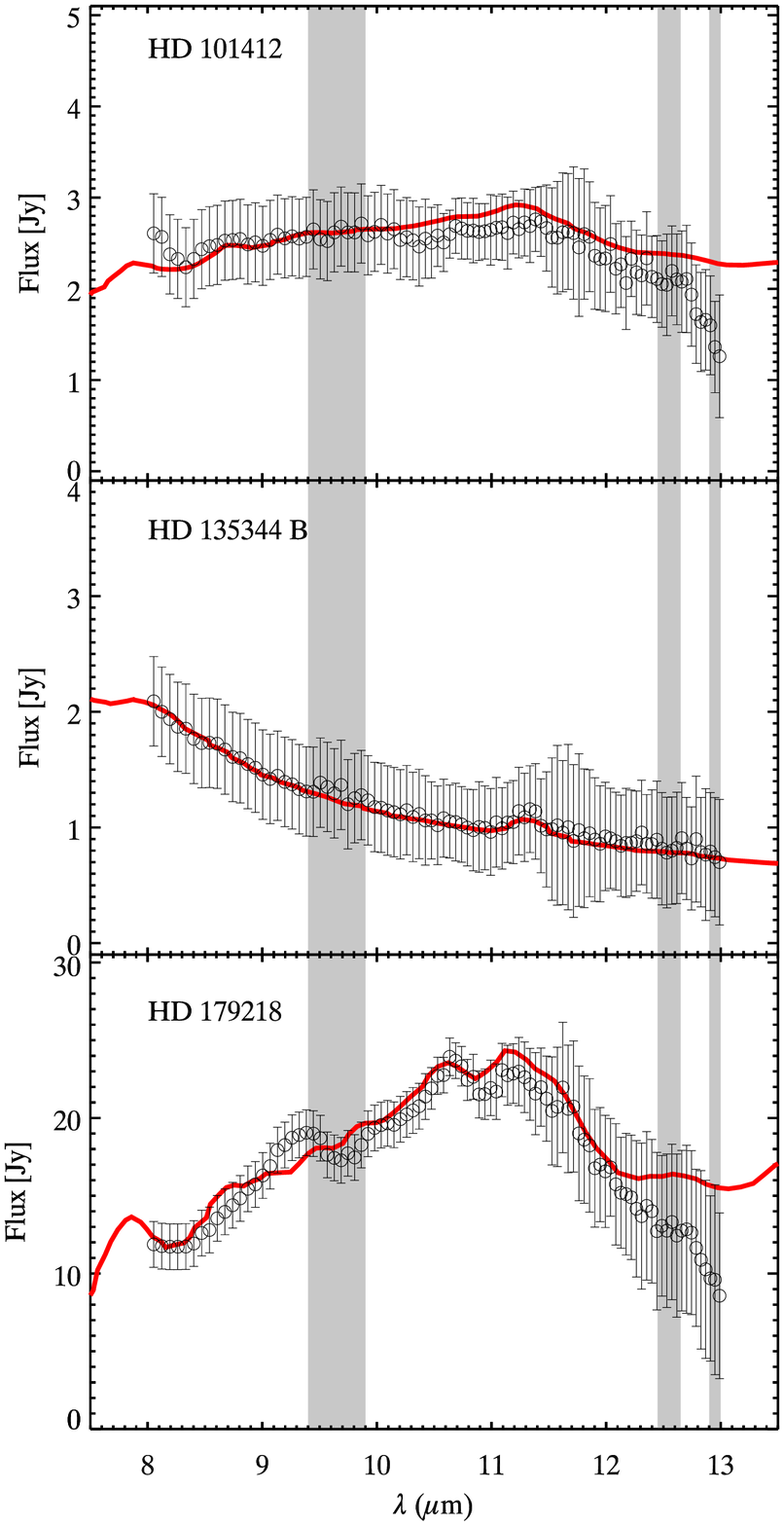}
   \caption{MIDI spectra (diamonds) of HD 101412, HD 135344 B and HD
     179218. The MIDI spectra were scaled to match the Spitzer IRS spectra (red
     line) at 9~$\mu$m. }
   \label{fig:calflux}
 \end{figure}

  The averaged MIDI spectra of the three targets are shown in
  Fig. \ref{fig:calflux}. The Spitzer spectrum is plotted for comparison. Each
  spectrum was scaled in order to match the Spitzer spectrum at 9~$\mu$m. MIDI
  and Spitzer spectra agree well. A systematic deviation of the MIDI spectrum
  from the Spitzer one is visible at the red edge of the N-band. This is due
  to a non-perfect correction of the MIDI spectrum for atmospheric
  absorption. 

  \subsection{Note on the data reduction of HD 135344 B}
  The data reduction process of HD 135344 B deserves a more detailed
  description. The brightness of this target is close to the sensitivity limit
  of the instrument. To check the quality of the data we used both the
  coherent (EWS) and incoherent (MIA) method.  Given the faintness of the
  source, we used a higher threshold for the ``good scans'' in MIA (60 -- 70
  \%). In all cases, small differences at the level of 10 -- 20 \% in the
  calibrated visibility are present. In particular, on the night of 2006 July
  12, EWS yields a lower visibility at short wavelengths. After different
  checks, the problem seems to be the fixed mask used to extract the signals
  with EWS. Both the bi-dimensional interferometric and photometric signal are
  slightly offset compared to the position of the EWS mask. For this reason we
  preferred to adopt the mask computed by MIA. With this procedure the two
  data reduction packages produce very similar results which are consistent
  with a precision of $\lesssim$ 10\%. 

  \section{Data analysis}
  The quantity measured by an interferometer is the Fourier transform of the
  brightness distribution, i.e. the complex visibility. Due to atmospheric
  turbulence, however, the phase of this complex number is corrupted and
  cannot be retrieved. The amplitude of the complex visibility -- the
  visibility -- can be measured and is compared to the corresponding quantity
  of theoretical brightness distributions. The visibility is a two-dimensional
  function in the so-called $u,v$-plane, where $u$ and $v$ are the spatial
  frequencies $B/\lambda$ ($B$ is the baseline length between the two
  telescopes, and $\lambda$ the wavelength) in two perpendicular
  directions. In the case of VLTI/MIDI, a single observation covers a range of
  spatial frequencies due to the spectral capability of the beam
  combiner. However, one should keep in mind that the brightness distribution
  of the target may vary from wavelength to wavelength.
  
  \smallskip
  \noindent
  
  Figs. \ref{fig:vis1}, \ref{fig:vis2} \& \ref{fig:vis3} show the calibrated
  MIDI visibilities of HD 101412, HD 135344 B and HD 179218 respectively. We
  plot the visibility versus spatial frequency rather than wavelength, in
  order to make a direct comparison with the spatial distribution of the
  [\ion{O}{i}] emission measured at 6300\AA. For convenience, we plot the MIDI
  wavelength scale at the top of the figure.  The slope of the visibility
  curve in the $N$-band depends on: 1) the increase in spatial resolution with
  decreasing wavelength and 2) the wavelength-dependence of the brightness
  distribution.
  
  The [\ion{O}{i}] visibility curves (dashed line) are calculated from the
  radial intensity profile presented in paper I and are discussed later in the
  text. The comparison of MIDI observations and [\ion{O}{i}] visibility is
  presented in Sec. \ref{sec:comp}. In this section we use a simple
  geometrical model aimed at deriving the geometrical properties (inner and
  outer radius, inclination and position angle) of the (mid-infrared) dust
  emitting region in the three protoplanetary disks.
  
  \subsection{Uniform ring model}
  In a first approximation the brightness distribution of a protoplanetary
  disk might be described by a uniform bright ring (UR). The formalism of the
  UR model is described in Appendix B. We emphasize that the UR model is a
  simple geometrical approximation of the dust emitting region. This model
  does not take into account the physical properties of the circumstellar disk
  (e.g. dust sublimation radius, disk outer radius, temperature gradient)
  or the dust properties (e.g. grain size). Such a model has been extensively
  used as an analytic model for near-infrared interferometric observations of
  disks around young stars (e.g. Millan-Gabet et al. \cite{millan}, Eisner et
  al. \cite{eisner04}, \cite{eisner07}) as well as mid-infrared
  interferometric data of planetary nebulae (Chesneau et
  al. \cite{chesneau}). The MIDI visibilities were fitted using an inclined
  uniform ring (UR) model.
  
  \smallskip
  \noindent
  
  We have searched for the best set of model parameters by minimizing the
  $\chi^2$ of the visibility measurements. The values of the best fit
  parameters and associated errors were computed using a Monte Carlo
  simulation. Assuming a normal error distribution of the measured visibility,
  we simulated 100 random data-sets around the observed values. The best fit
  parameters and errors correspond to the mean and the standard deviation of
  the 100 fits.

  \subsection{HD 101412}
\begin{figure}
    \centering
    \includegraphics[width=8cm]{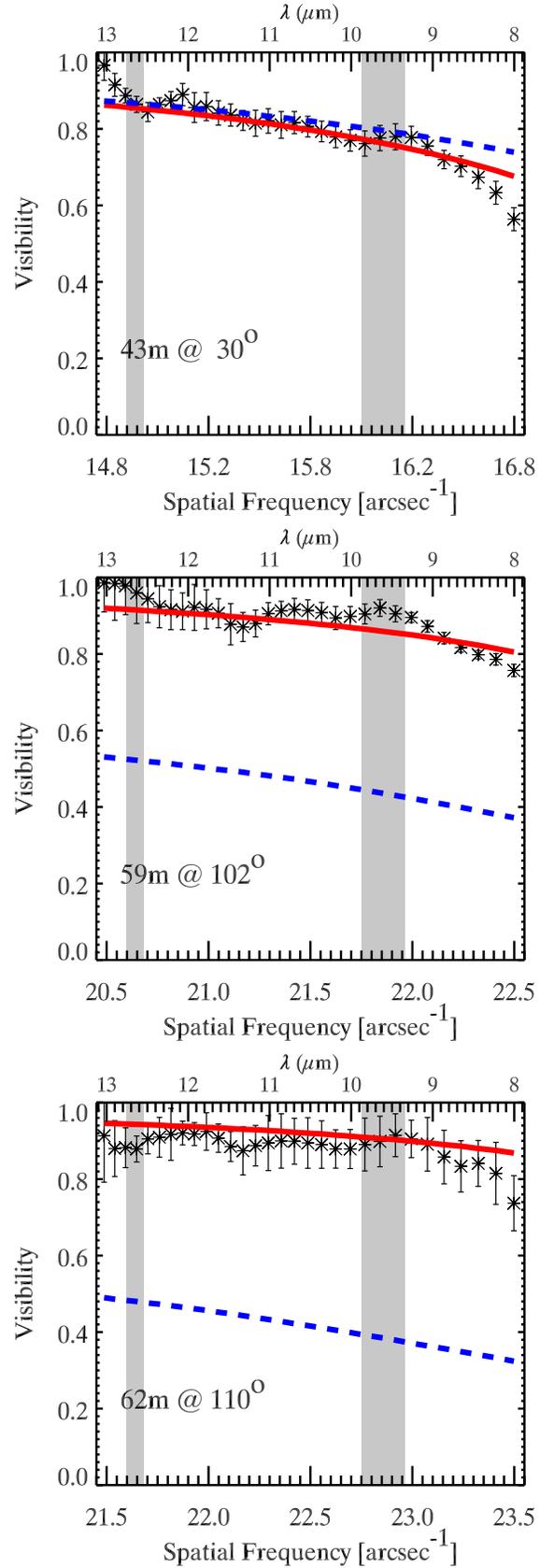}
    \caption{Calibrated visibility amplitude at different baselines for HD
      101412. The dashed line in the visibility panel represents the
      visibility computed from the intensity versus radius profile of the
      [\ion{O}{i}]6300\AA~ emission line. The solid line is the best fit of
      the inclined UR model. Visibility errors are computed as standard
      deviations resulting from calibrations with different calibrator stars.}
    \label{fig:vis1}
  \end{figure}

  HD 101412 (Fig. \ref{fig:vis1}) is barely resolved at all epochs. Along 
  the three baseline position angles the visibility is lower at short
  wavelengths and increases from 8~$\mu$m to 9~$\mu$m -- 10~$\mu$m. Differences
  in the absolute value and shape in the three measurements may suggest an
  asymmetric emission. In particular, HD 101412 is more resolved (i.e. shows a
  more extended emission) with the shortest baseline, 43 m at 30\degr. Along
  this direction the visibility goes from 0.55 at 8~$\mu$m to 0.9 at
  13~$\mu$m. Around 11.3~$\mu$m the disk appears larger than at adjacent
  wavelengths (with a small dip in visibility). This coincides with PAH
  emission. The high visibility values suggest a compact (mid-infrared) dust
  emission. The decreasing visibility with decreasing wavelength is due to the
  increase in spatial resolution. The three visibility measurements of HD
  101412 are best fitted by a UR model (Fig. \ref{fig:vis1}, solid line) of
  the inner and outer radius of 0.4 and 1.9 AU, an inclination of
  80$\degr$ and position angle of the disk major axis of 38$\degr$. The almost
  edge-on orientation found here explains why the [\ion{O}{i}] line profile
  shows structures that are not seen in other [\ion{O}{i}] profiles of Herbig
  stars; thanks to the high inclination, the projected velocities are close to
  the real velocities, and the Doppler-induced spread in the spectrum is
  optimal.

  \smallskip
  \noindent
  
  According to the classification of Meeus et al. (\cite{meeus}), the SED of
  HD 101412 suggests a self-shadowed disk geometry. In this case the
  mid-infrared disk emission is expected to arise only from the inner rim of
  the disk. This is consistent with the small size measured with MIDI.
  
  \begin{figure}[!h]
    \centering
    \includegraphics[width=8cm]{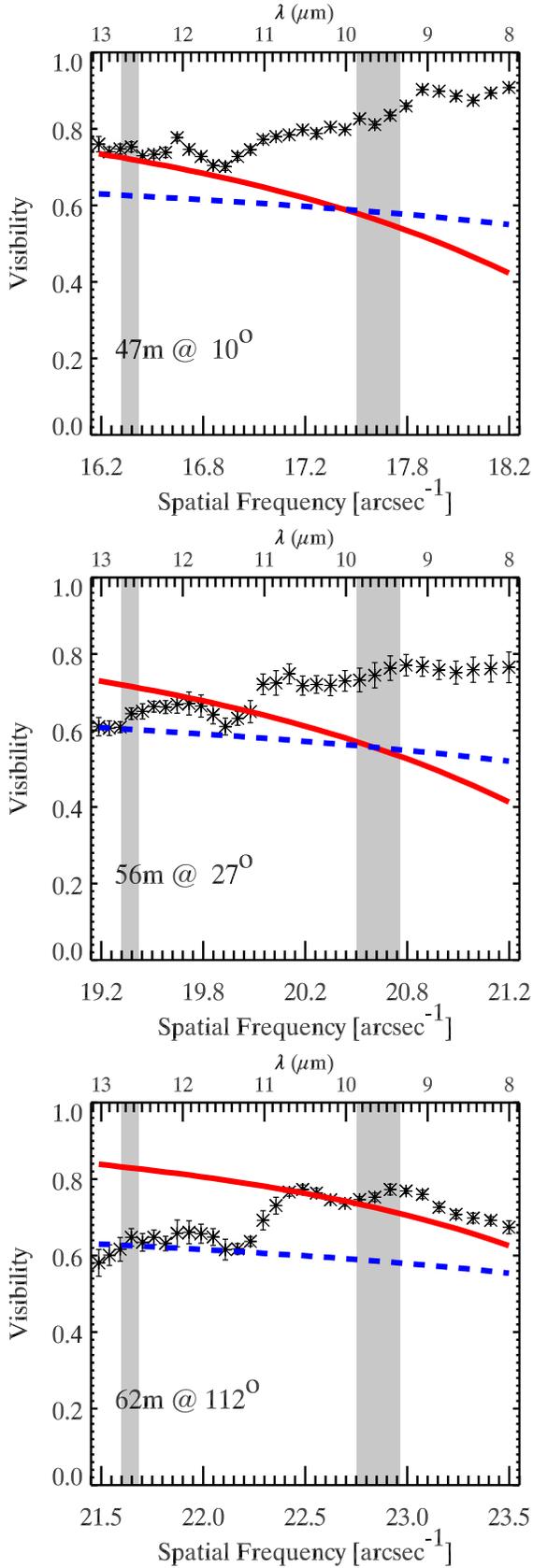}
    \caption{Same as Fig. \ref{fig:vis1} for HD 135344 B.}
    \label{fig:vis2}
  \end{figure}
 
  \begin{figure}[!h]
    \centering
    \includegraphics[width=9cm]{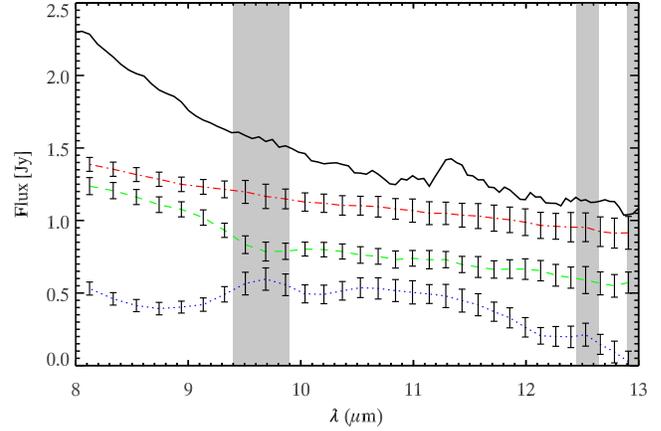}
    \caption{Correlated spectrum of HD 135344 B measured on 2006 May 13
      (dotted), 2006 June 13 (dashed) and 2006 July 12 (dot-dashed). The
      spectra were shifted for clarity. The mean, single telescope, spectrum
      is also plotted (solid line). The absence of the PAH emission feature at
      11.3~$\mu$m in the correlated spectrum is a signature of spatially
      extended PAH emission compared to the continuum emission.}
    \label{fig:CF}
  \end{figure}

  \subsection{HD 135344B}
  HD 135344 B is resolved with all baselines (Fig. \ref{fig:vis2}). Along the
  47 m baseline (position angle 10\degr) the visibility decreases with
  wavelength.  Along the second and third baselines, 56 m and 62 m
  respectively, the visibility shows a similar shape: nearly constant between
  8~$\mu$m -- 11~$\mu$m with a small bump around 9.2 -- 9.4~$\mu$m, a drop at
  11.3~$\mu$m and almost constant at longer wavelength. Despite the higher
  resolution, the source is less resolved at short wavelengths. This means
  that at longer wavelengths the target is significantly more extended due to
  emission of the colder outer disk which only starts to show up at these
  wavelength. This specific pattern is sometimes observed in mid-infrared
  interferometric observations of young PMS stars (e.g. V1647 Orionis,
  Abraham et al. \cite{abraham}; FU Orionis, Quanz et
  al. \cite{quanz}). Reasonable fitting of this visibility shape is obtained
  with a (physical) Keplerian, flared disk model where temperature and surface
  density are prescribed by a broken power law distribution ($T \propto
  r^{-q}$; $\Sigma \propto r^{-p}$), i.e. the disk's surface is cooler at
  larger radii. The result of the UR model (Fig. \ref{fig:vis2}, solid line
  and Table \ref{tab:ring}) are: inner radius 0.05 AU, outer radius 1.8 AU
  (assuming a distance of 140 pc van Boekel et al. \cite{boekel05}),
  inclination 60 $\degr$ and position angle of the disk major axis
  180\degr. The inclination found here is slightly higher than the result of
  Doucet et al. (\cite{doucet}) (45$\degr$), obtained with direct imaging at
  20.5~$\mu$m. Dent et al. (\cite{dent}), modeling the J=3-2 transition of
  $^{12}$CO, find a smaller disk inclination (11$\degr$, disk nearly
  face-on). Nevertheless, the UR model is not able to reproduce the decreasing
  visibility observed with MIDI and the disk parameters are not well
  constrained.

  We converted the MIDI visibility to the Gaussian full width half maximum
  (FWHM) at each spatial frequency. The FWHM represents the extent of the
  emitting region and is defined as:

  \begin{equation}
    \theta = \sqrt{\frac{ln(V(sf))}{-3.56 sf^2}}
  \end{equation}
  
  where $V(sf)$ is the MIDI visibility at spatial frequency $sf$. A Gaussian
  distribution is only a crude approximation of the real brightness
  distribution valid at high visibilities. Similarly to Quanz et
  al. (\cite{quanz}), we also fitted the FWHM measured on each baseline with
  an ellipse wavelength by wavelength. The best-fit ellipse parameters at
  three reference wavelengths (9~$\mu$m, 11~$\mu$m, 12.5~$\mu$m) are listed in
  Table \ref{tab:ellipse}. As expected, the disk looks larger at longer
  wavelengths. The ellipse semi-major axis increases from 1.1 AU at 9~$\mu$m
  to 1.9 AU at 12.5~$\mu$m. The inclination and position angle varies by $\sim$
  5\degr ~and $\sim$ ~9\degr respectively from 9~$\mu$m to 11~$\mu$m. The
  11~$\mu$m and 12.5~$\mu$m ellipses have similar (within $\sim$ 3\degr)
  inclinations and position angles. The geometry of the mid-infrared dust
  emission derived with the UR model and the ellipse fitting are consistent
  with each other.

  \smallskip
  \noindent
  
   \begin{figure*}[!t]
    \centering
    \includegraphics[width=16cm]{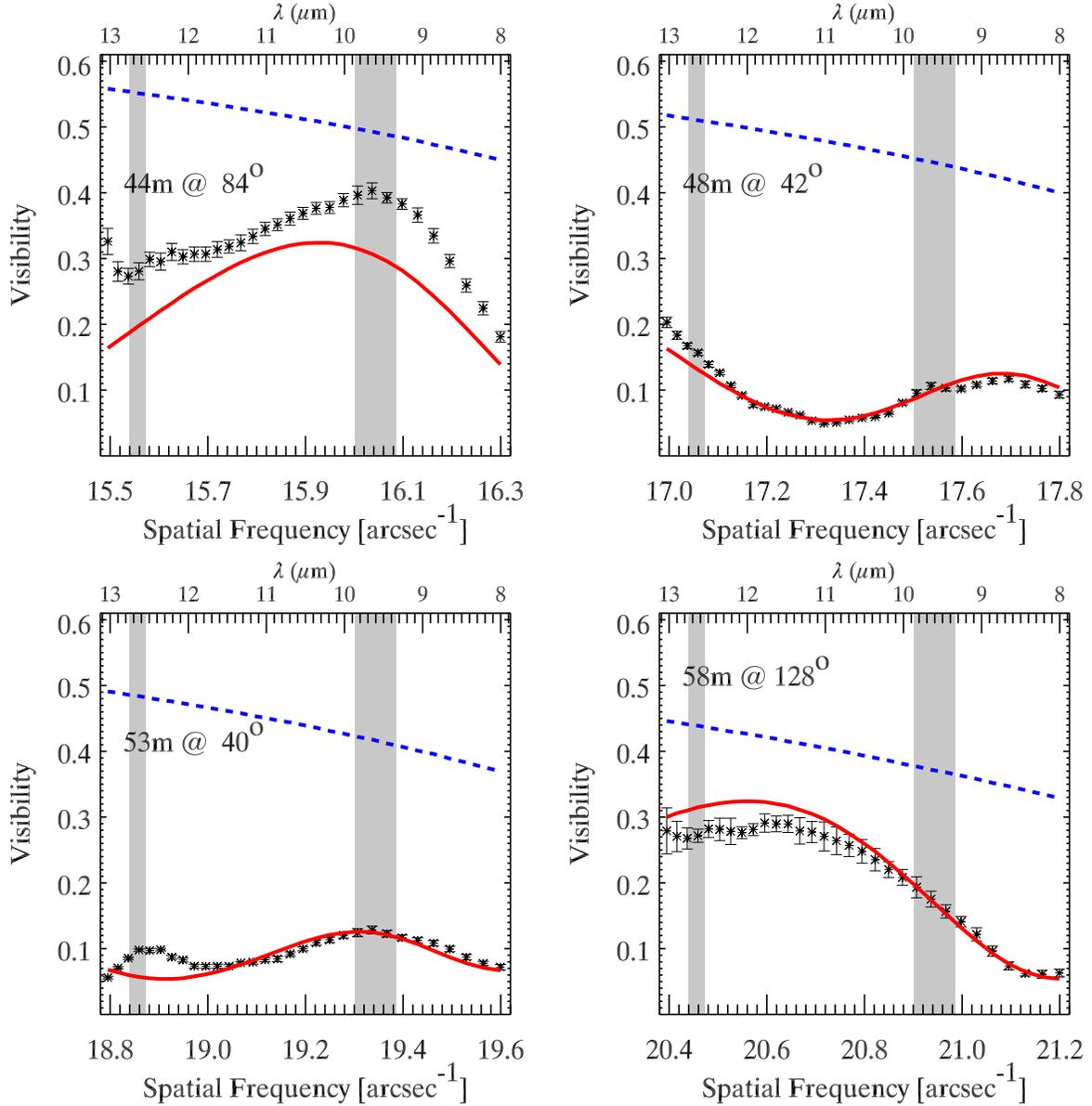}
    \caption{Same as Fig. \ref{fig:vis1} for HD 179218. The continuous
      line here represents the best fit two-component disk model.}
    \label{fig:vis3}
  \end{figure*}

   The SED of HD 135344 B is characterized by a relatively low infrared excess
  over the stellar photosphere up to 13~$\mu$m and by a strong infrared excess
  beyond 14~$\mu$m. Moreover, the mid-infrared spectrum
  (Fig. \ref{fig:calflux}) lacks silicate features and only weak PAH emission
  is detected. Such an SED is peculiar among group I sources and suggests a
  more complex disk structure. The missing near- to mid-infrared excess and
  the fast increase at longer wavelengths have been attributed to the presence
  of a large dust-free gap by Brown et al. (\cite{brown}). They proposed a
  so-called  ``cold disk'' structure for the material around HD 135344 B. In
  their model the disk is devoid of dust between 0.45 AU -- 45 AU (in
  radius). According to Brown et al., warm dust is present in an inner ring
  between  0.18 AU -- 0.45 AU. The MIDI observations presented here are
  consistent with this model. The Gaussian FWHM, if scaled to the distance
  adopted by Brown et al., yields an emitting region between 0.4 -- 0.7
  AU. Although these radii are larger than those found by Brown et al. (note
  that we used a simple Gaussian model), the two estimates agree within the
  uncertainty of the data and the analysis.
  
  \smallskip
  \noindent
  
  Because MIDI is only sensitive to dust emitting at 10~$\mu$m, our data set
  does not allow us to claim the presence or absence of cold matter beyond the
  inner warm disk. The shadow of the inner disk may obscure a large portion of
  the disk which re-emerges only at large radii. Both the SED and MIDI
  observations are only sensitive to the dust emitting in the mid-infrared,
  both the SED and our MIDI observations are not able to distinguish between a
  (partially) self-shadowed disk and a disk with a gap.

  \smallskip
  \noindent

  The correlated spectrum of HD 135344 B for the different measurements is
  plotted in Fig. \ref{fig:CF}. At all epochs the correlated spectrum mimics
  the total (single-telescope) spectrum, although deviations are visible. The
  11.3~$\mu$m PAH emission detected in the total spectrum is not present in
  any of the correlated spectra which indicates that the PAH 11.3~$\mu$m
  emission region is completely resolved at the angular resolution of our
  measurements.

  \subsection{HD 179218}
  Despite being the most distant object of our sample, HD 179218 is the most
  spatially resolved source of the three objects. The visibility varies with
  the position angle of the baseline and is always below 0.4. Along the first
  baselines the visibility increases rapidly from 8 to 9~$\mu$m and decreases
  afterwards. Along the second and third baselines (48 and 53 m), the
  visibility is much lower and shows a characteristic sinusoidal
  modulation. In the last case, the visibility increases from a very low value,
  0.05 at 8~$\mu$m to 0.27 at 11.7~$\mu$m and decreases afterwards.
 
  A sinusoidal pattern of visibility is produced if the brightness
  distribution of the target displays sharp edges, e.g. in the case of a ring,
  uniform disk or a binary with two unresolved components. A smooth emission
  region, such as a Gaussian disk, would not produce such variations. We can
  immediately exclude the binary hypothesis: the MIDI observations of HD
  179218 cover a range in position angles over 90$\degr$ and the visibility
  always shows a sinusoidal pattern. In the case of a binary, the pattern is PA
  dependent and completely disappears when the PA of the baseline and that of
  the binary axis are aligned. We attempted to fit the MIDI visibility with
  the UR model but although this model is able to reproduce the sinusoidal
  modulation of the visibility, it fails to reproduce the visibility at the
  position of the local ``minima''. The minimum in the ring
  model is zero while the observed visibility reaches a minimum value of $V =
  0.05$. If these were ``true'' zeroes of the visibility, i.e. the location
  between two adjacent lobes of the visibility curve, we would have measured a
  contrast of $V \sim 0.0076$\footnote{This value, given the finite band-width
    of the MIDI wavelength bins, corresponds to the visibility null reachable
    by the interferometer at low spectral resolution (PRISM).}. We then
  attempted to fit the data with a two-component model. The first component
  represents the disk emission at small radii and the second component
  represents the disk surface at larger radii. Our simulations showed that the
  first component is only barely resolved. We used a uniform ring for both the
  internal and external component. Using a different brightness distribution
  for the internal component does not improve the model fit. The
  fractional flux contribution of the first component, $f_{int}$, was left as
  a free parameter in the fit. For simplicity, $f_{int}$, was assumed to be
  constant at all wavelengths. The best fit parameters are listed in Table
  \ref{tab:HD179218}. Inclination and position angle are in agreement with
  those found with the single ring model ($i$ = 57$\degr$, PA =
  198$\degr$). The inner component contributes 20\% of the total mid-infrared
  flux. According to this model, the disk around HD 179218 is either truncated
  or obscured by its own shadow between 3.2 and 14.5 AU radii from the star
  (Table \ref{tab:HD179218}).

  Previous infrared interferometric observations of HD 179218 have been
  carried out by by Leinert et  al. (\cite{leinert04}) and Liu et
  al. (\cite{liu}) in the mid-infrared and by Monnier et al. (\cite{monnier})
  in the H-band. Leinert et al. also measured low visibilities with an 80
  meter baseline. Monnier et al. (\cite{monnier}) measured a squared
  $H$-band visibility of the order of 40\%, corresponding to a Gaussian FWHM
  of 3.8 AU. This is in good agreement with our two-component model, where the
  inner component extends up to 3.5 AU from the star. Liu et al. modeled MMT
  nulling interferometric data by means of a ring of diameter 27 $\pm$ 5
  AU. This large ring is, within the uncertainties, consistent with the outer ring
  of our two-component model.

  \smallskip
  \noindent
  
  The agreement between the MIDI data and the simple geometrical model
  proposed here is quite good. However, deviations from the model are clearly
  visible. Similarly to HD 135344 B, our observations are not able to
  distinguish between a real gap (i.e. a region free of material) and a
  shadowed region.

  \subsubsection{Differential phase}
\begin{figure}
    \centering
    \includegraphics[width=9cm]{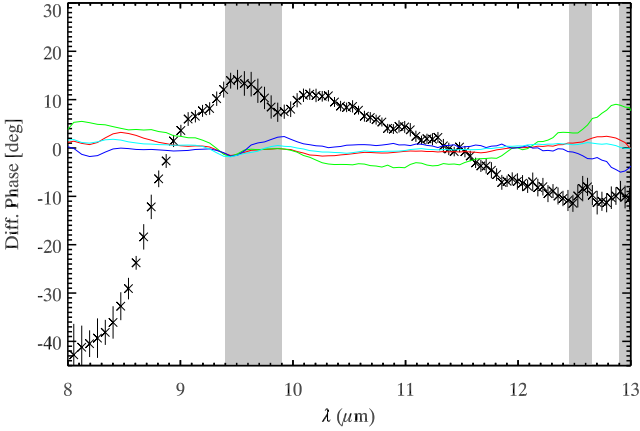}
    \caption{Differential phase of HD 179218 measured on 2006 May 16. A 55$^\circ$
      phase variation is detected at short wavelengths. The four continuous
      lines are the differential phases for four different calibrators
      observed during the same night. The plotted phases are not corrected for
      the effect of water vapor in the delay lines of the VLTI. See Sec. 4.4.1
      for details.}
    \label{fig:DP}
  \end{figure}

  As a byproduct of the EWS data reduction software, the relative phase
  between the different spectral channels of MIDI can be retrieved. This
  so-called differential phase is the real differential phase of the target on
  the sky, modulated by the atmosphere. The largest effect of the latter is an
  offset of the differential phases, which is linear with respect to
  wavenumber ($\propto \lambda^{-1}$). The EWS software subtracts the best-fit
  linear component of the measured differential phase, as this information is
  lost anyway. Secondary atmospheric effects remain: the two light beams,
  coming from the two telescopes, pass through different amounts of air and
  water vapor before they reach the MIDI beam combiner. However, these smooth
  variations in the measured differential phase are of the order of a few
  degrees. For all targets the differential phase is nearly constant with
  wavelength and centered at zero degrees. The only exception is the
  measurements of HD 179218 on 2006 May 16 (Fig. \ref{fig:DP}). Along this
  baseline, a large phase variation is detected at short wavelengths. The
  phase varies by 55$\degr$ between 8 and 9~$\mu$m. A flip of 180$\degr$ is
  expected when the visibility amplitude crosses a null, i.e. when the spatial
  frequencies covered by the observations are located between two adjacent
  lobes of the visibility curve. In the case of HD 179218 the phase variation
  is smooth with lower amplitude. The phase variation is then ascribed to a
  shift of the photo-center of the emitting source. This suggests a deviation
  from centro-symmetric emission along this direction.

  The differential phase plotted in Fig. \ref{fig:DP} was not calibrated for
  the effect of water vapor that is present in the delay line tunnel of the
  VLT interferometer. The two light beams most likely pass through a different
  amount of air and water vapor before reaching the MIDI beam combiner. This
  difference introduces an instrumental differential phase. Fig. \ref{fig:DP}
  also shows the differential phase measured with four different calibrators
  observed during the same night with the same acquisition mode (SCI
  PHOT). Since calibrators are (usually) point-like their differential phase
  should be zero. The deviation from zero degree in this case is produced by
  the water vapor. The effect of water vapor is present, but is very small
  compared to the phase variation observed toward HD 179218.

  The variation in the differential phase along the 58 m baseline is caused by
  asymmetric emission. This is in agreement with the inclination of the disk
  (57$\degr$) and its flared structure. Due to the flaring, the far side of
  the disk emits more light in the direction of the observer. This means that
  the photo-center of this emission (described here by the ring) is no longer
  coincident with the photo-center of the almost unresolved inner disk. It is
  difficult to derive a quantitative estimate for this offset, mainly due to
  the loss of the linear differential phase information. Estimating from the
  magnitude of the differential phase variation ($\sim$60$\degr$), it must be
  around 4$\pm$2~mas.

  \section{Comparison to the optical gas emitting region}\label{sec:comp}
\begin{figure}
    \centering
    \includegraphics[width=9cm]{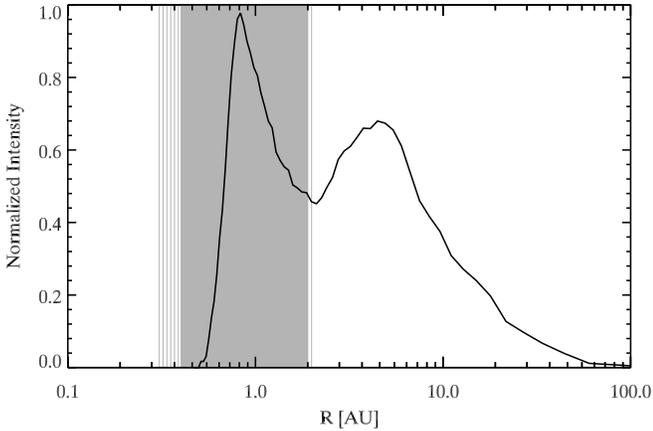}
    \caption{Normalized [\ion{O}{i}] 6300\AA\ intensity profile for HD
      101412. $I_{[\ion{O}{i}]}(R) \times S(R)$ is the total [\ion{O}{i}] line
      luminosity in a ring at distance $R$ from the star. This was obtained
      averaging the results of observation $B1$ and $B2$ of paper I. In the
      specific case of HD 101412, $I_{[\ion{O}{i}]}(R)$ was recomputed with an
      inclination of 80\degr. The filled region represents the position of the
      dust emission region as determined by MIDI. The striped regions indicate
      the uncertainty on the latter. }
    \label{fig:OI_1}
  \end{figure}

  In paper I very high resolution optical spectra of HD 101412, HD 134355B and
  HD 179218 were presented. In all three targets, we spectrally resolve the
  [\ion{O}{i}] 6300\AA~ emission line. Assuming that the disk is in Keplerian
  rotation, the [\ion{O}{i}] emission line profile can be translated into a
  radial profile, i.e. the intensity-versus-radius curve of the oxygen line
  emission, $I_{[\ion{O}{i}]}(R)$. Figs. \ref{fig:OI_1}, \ref{fig:OI_2} and
  \ref{fig:OI_3} show the normalized [\ion{O}{i}] radial profile derived in
  paper I: $I_{[\ion{O}{i}]}(R) \times S(R)$, where $S(R) = \pi (R_{out}^2 -
  R_{in}^2)$ and $R_{out}$ and $R_{in}$ outer and inner radius of the ring at
  radius $R$. $I_{[\ion{O}{i}]}(R) \times S(R)$ is the total luminosity in a
  ring at distance $R$ from the central star. The [\ion{O}{i}] radial profile
  shown here is based on the average of the spectra presented in Paper I. In
  the case of HD 101412, $I_{[\ion{O}{i}]}(R)$ was recomputed using a disk
  inclination of 80\degr ~to be consistent with the result of Sec. 5. Using a
  larger inclination has the effect of shifting the emission further away from
  the star as can be seen comparing Fig. \ref{fig:OI_1} with Fig. 11 of paper
  I. The filled region represents the position of the dust emission region as
  determined by MIDI. The striped regions indicate the uncertainty on the
  latter.
  
  \smallskip
  \noindent
  
  In order to compare the [\ion{O}{i}] result with the dust one we need to
  provide a visibility model for the MIDI observations. From
  $I_{[\ion{O}{i}]}(R)$, it is possible to derive a brightness distribution
  ($I_{[\ion{O}{i}]}(\alpha,\beta)$, with $\alpha$ and $\beta$ the angular
  distance from the central star) assuming that the line emission from the
  disk surface is axis symmetric. Using $I_{[\ion{O}{i}]}(\alpha,\beta)$ as
  brightness distribution, the visibility model ($V_{[\ion{O}{i}]}(u,v)$) was
  computed. The expected visibility at the spatial frequencies mapped with
  MIDI was computed interpolating $V_{[\ion{O}{i}]}(u,v)$ at the corresponding
  positions in the uv-plane. The dashed line in Figs. \ref{fig:vis1},
  \ref{fig:vis2} \& \ref{fig:vis3} is $V_{[\ion{O}{i}]}(u,v)$ at the spatial
  frequency covered by MIDI observations.  

  \begin{table}
    \caption{Best fit parameters of the UR model for HD 101412 and HD 135344
      B. $R_{in}$ and $R_{out}$ are the inner and outer ring radii and are
      given in  milliarcsec (first row) and AU (second row). The last column
      lists the value of the reduced $\chi^2$. Values close to one indicate a
      good agreement between model and observations. }
    \label{tab:ring}
    \centering
    \begin{tabular}{llllll}
      \hline\hline
      Star      & $R_{in}$    & $R_{out}$    & i  & $\phi$  & $\tilde{\chi}^2$\\
                & [mas/AU]  & [mas/AU]       & [$\degr$] & [$\degr$] & \\
      \hline
      HD 101412 & 2.5 $\pm$ 0.6  & 11.9 $\pm$ 0.6  & 80 $\pm$ 7 & 38 $\pm$ 5  & 0.4  \\
                & 0.4 $\pm$ 0.1  &  1.9 $\pm$ 0.1  & & & \\     
      HD 135344 & 0.35 $\pm$ 1.79 &  12.8 $\pm$ 1.4  & 60 $\pm$ 10&  180 $\pm$ 60 & 3.9  \\
                & 0.05 $\pm$ 0.25 &  1.8 $\pm$ 0.2  & & & \\
       \hline\hline
    \end{tabular}
  \end{table}

  \begin{table}
    \centering
    \caption{Best-fit ellipse parameters for HD 135344 B. The ellipse is
      fitted to the FWHM measured on the three baselines. Ellipse axis are
      computed assuming a distance of 140 pc.}\label{tab:ellipse}
    \begin{tabular}{llll}
      \hline\hline
      Wavelength            & 9~$\mu$m & 11~$\mu$m & 12.5~$\mu$m \\
      \hline
      Semimajor axis [AU]   & 1.1  & 1.5  & 1.9  \\
      Semiminor axis [AU]   & 0.6  & 0.8  & 0.9  \\
      Inclination [\degr]   & 53   & 58   & 61   \\
      Position angle [\degr]& 171  & 163  & 163  \\
      \hline\hline
    \end{tabular}
  \end{table}

  \begin{table}
    \caption{Best fit parameters and standard deviation of the two-component
      disk model for HD 179218. Radii are given in AU assuming a distance of
      240 pc. The inclination and position angle of the two components were
      assumed to be equal for both structures. $f_{in}$ is the fractional
      contribution of the inner component to the total flux in the
      mid-infrared.}
    \label{tab:HD179218}
    \centering
    \begin{tabular}{ll}
      \hline\hline
      HD 179218   & \\
      \hline
      1$^{st}$ component: $R_{in}$  (AU)  & 0.3 $\pm$ 0.1 \\
      1$^{st}$ component: $R_{out}$ (AU)  & 3.2 $\pm$ 0.1 \\
      2$^{nd}$ component: $R_{in}$  (AU)  & 14.5 $\pm$ 0.5 \\
      2$^{nd}$ component: $R_{out}$ (AU)  & 22.6 $\pm$ 0.4 \\
      inclination (\degr) & 57 $\pm$ 2 \\
      PA         (\degr)& 23 $\pm$ 3 \\
      $f_{in}$   & 0.2 $\pm$ 0.02 \\
      \hline\hline
    \end{tabular}
  \end{table}

  \subsection{HD 101412}
  As shown in Fig. \ref{fig:vis1}, {\bf $V_{[\ion{O}{i}]}(u,v)$} is similar to
  the observed visibility along the first baseline (43 m, 30$\degr$) and is
  clearly lower than along the other two directions. This is clearer when
  looking at Fig. \ref{fig:OI_1}. The dust emission is located within $\sim$ 2
  AU, while the gas emission appears more extended with a bright peak at
  $\sim$ 5 -- 6 AU. In terms of radial extent, this means that the
  [\ion{O}{i}] emission extends to larger radii than the mid-infrared
  emission. Beyond the first peak the [\ion{O}{i}] intensity drops off as one
  would expect for a self-shadowed disk.

  \subsection{HD 135344 B}
\begin{figure}
    \centering
    \includegraphics[width=9cm]{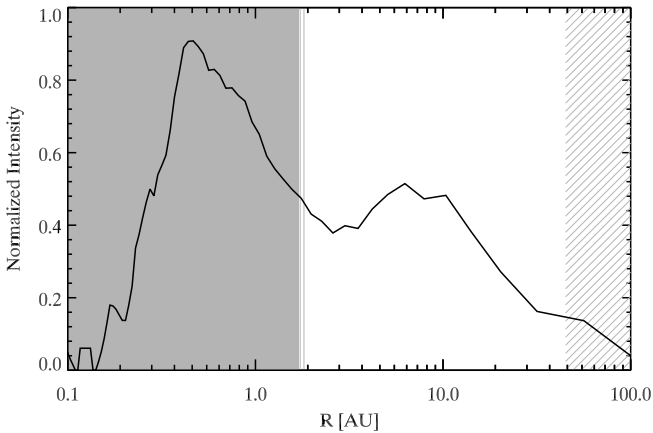}
    \caption{Same as Fig. \ref{fig:OI_1} for HD 135344 B. The diagonal striped
      region represents the cold dust in a flared geometry detected by Doucet
      et al. (\cite{doucet}) and Brown et al. (\cite{brown}). This cannot be
      seen with MIDI because it radiates at wavelengths longer than those
      traced with MIDI.}
    \label{fig:OI_2}
  \end{figure}

  Similarly to HD 101412 the [\ion{O}{i}] brightness profile of HD 135344 B
  shows a double peak with a first stronger peak at small radii ($\sim$ 0.5
  AU) and a second fainter peak at $\sim$ 5-10 AU (Fig. \ref{fig:OI_2}). On
  the other hand, the 10 $\mu$m dust emission seems to come from the inner 1 -
  2 AU of the disk surface. However, the [\ion{O}{i}] result suffers from
  overlapping photospheric absorption and thus have to be interpreted with
  more caution (paper I). The diagonal striped region in Fig. \ref{fig:OI_2}
  represents the cold dust detected at longer wavelengths by Doucet et
  al. (\cite{doucet}) and by Brown et al. (\cite{brown}).
 
  \subsection{HD 179218}
\begin{figure}
    \centering
    \includegraphics[width=9cm]{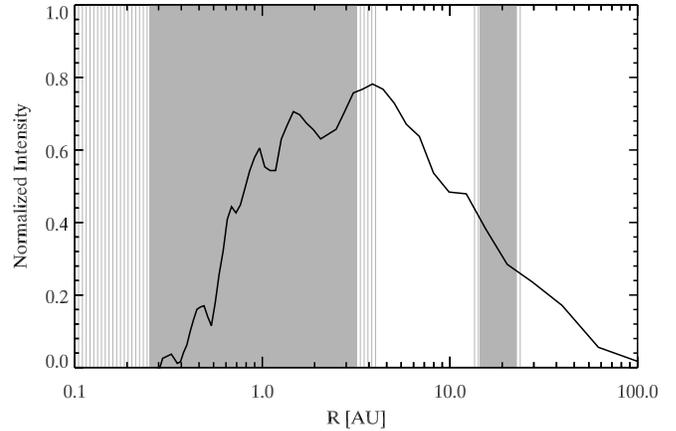}
    \caption{Same as Fig. \ref{fig:OI_1} for HD 179218.}
    \label{fig:OI_3}
  \end{figure}

  As shown in Fig. \ref{fig:vis3}, the [\ion{O}{i}] model is not able to
  reproduce the observed visibility of HD 179218. The observed visibility is
  in all cases lower than {\bf $V_{[\ion{O}{i}]}$}. Moreover, the oxygen model
  increases with wavelength while the observed visibility shows a sinusoidal
  variation. The [\ion{O}{i}] intensity peaks at 4 -- 6 AU
  (Fig. \ref{fig:OI_3}) and decreases outwards. According to our two-component
  disk model, emitting dust is present within the oxygen peak. A further
  contribution to the mid-infrared emission, 80\% in terms of flux, comes from
  a ring located between $\sim$ 15 -- 23 AU. At this distance, the
  [\ion{O}{i}] line intensity is reduced. Our result is consistent with a
  flared geometry for the disk around HD 179218; the inner, curved, rim is
  represented by the inner UR while the outer UR may represents the flared
  disk surface at larger radii. The two dust emission regions are separated by
  a ``dark'' (non emitting) region.
    
 \begin{figure}[!ht]
    \centering
    \includegraphics[scale=0.5]{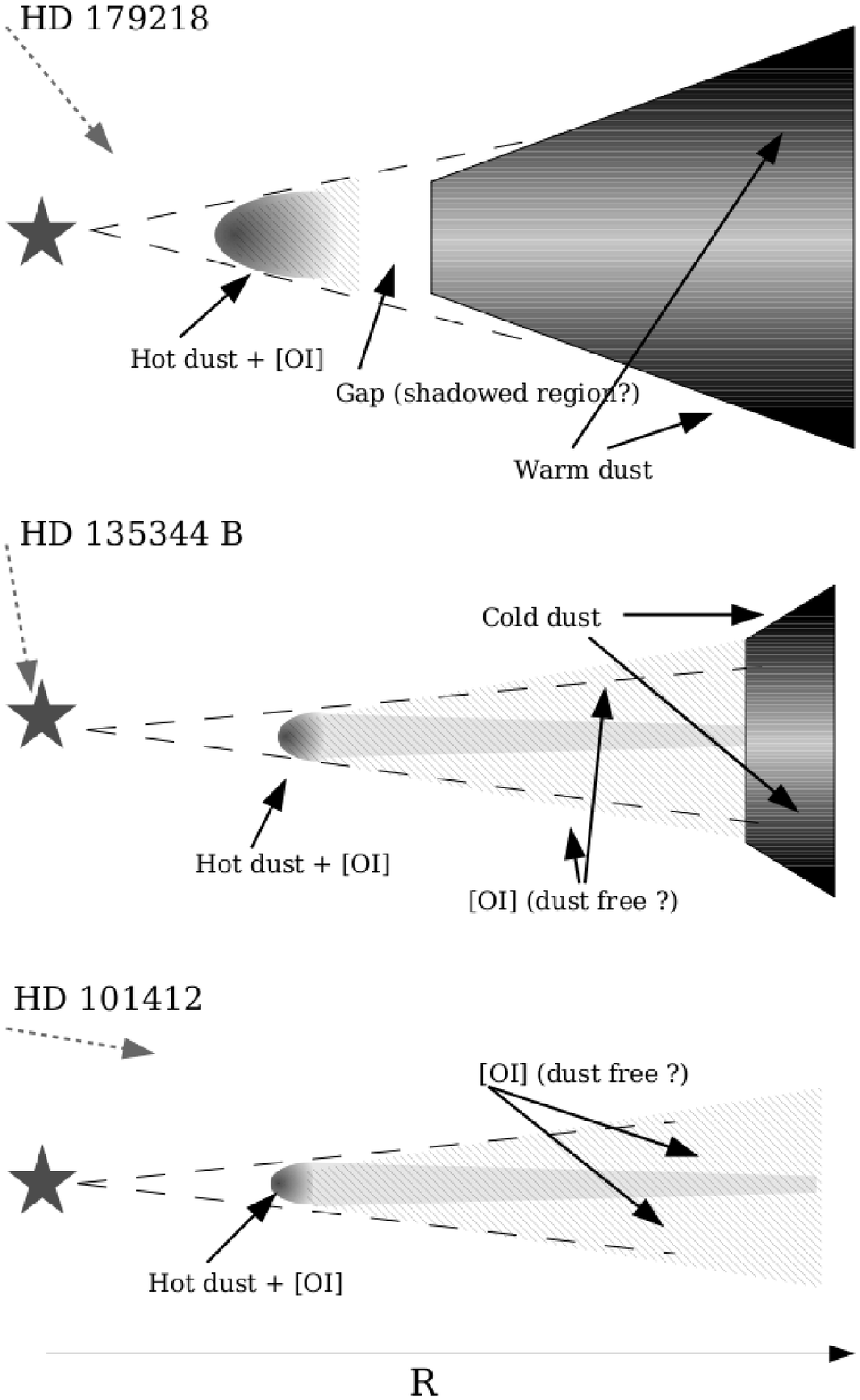}
    \caption{Simplified sketch of the disk around the three program stars. The
      regions where the dust and [\ion{O}{i}] emissions are expected to arise
      are indicated. The inclination of the disk to the line of sight is
      shown by the dashed arrow at the upper-left corner. {\it HD 179218}: the
      disk is large and flared. No dust emission is detected between 3 -- 15
      AU. Most of the silicate mid-infrared emission arises from the warm dust
      located in the surface layers of the outer part while the gas emission
      is mainly confined within 10 AU to the star. {\it HD 135344 B}: the 10
      $\mu$m emission arises from a compact region close to the star. Cold
      dust is detected at longer wavelengths. [\ion{O}{i}] 6300\AA ~emission is
      detected within the 10 $\mu$m emitting region and further away at 5 --
      10 AU where no dust emission is detected. {\it HD101412}: a compact dust
      ring that directly faces the star is responsible for the mid-infrared
      excess. The [\ion{O}{i}] 6300\AA ~line arises from the same
      region. Beyond this ring, the disk is shadowed. A dust-free flaring
      layer emerges from the shadow at larger radii. In all three cases, gas
      is likely present within the dust sublimation radius.}
    \label{fig:sketch}
  \end{figure} 

  \section{Discussion and conclusion}
  In this paper we presented the first direct comparison of gas and dust
  emission from the surface of three protoplanetary disks. The comparison of
  high optical spectroscopy with infrared interferometric observations gives
  some insight into the relative size and shape of the gas and the dust emitting
  region.
 
  \smallskip
  \noindent
  
  For HD 179218 the new MIDI results presented here and the UVES results of
  paper I are in good agreement with a flared disk structure. The 10 $\mu$m
  emission comes from two separate regions: an inner part located between
  $\sim$ 0.3 -- 3 AU (likely the disk inner rim) and an outer part between
  $\sim$ 15 -- 23 AU (flared disk surface). No dust emission is detected
  between 3 -- 15 AU. The [\ion{O}{i}] brightness profile shows instead a
  single strong peak centered at $\sim$ 3 -- 6 AU. The difference between gas
  and dust emitting regions for HD 179218 may reflects a real different
  structure for the two components or might be caused by other effects such
  as: 1) the action of chemical processes that reduce the abundance of OH in
  the outer part of the disk (in this case plenty of gas may still exist but
  we are not able to see it); 2) a contrast effect of the [\ion{O}{i}]
  emission. In the latter case the [\ion{O}{i}] emission is much stronger in
  the inner disk and it may outshine the [\ion{O}{i}] from the outer disk. The
  presence of a dust gap needs further investigation as it may indicate
  possible ongoing planet formation.

  \smallskip
  \noindent
  
  HD 135344 B and HD101412 show a compact 10 $\mu$m emitting region. The
  oxygen instead has a double peak brightness radial profile with the first
  peak coincident with the dust emission and a second peak further away from
  the star at $\sim$ 5 -- 10 AU. In the case of HD 135344 B extended dust
  emission is found at 20.5 $\mu$m (Doucet et al. \cite{doucet}). Modelling
  the SED of HD 135344 B, Brown et al. (\cite{brown}) propose a ``cold-disk''
  geometry where the disk is mostly void of dust between 0.4 -- 45 AU. Dust
  is present within the inner 0.4 AU and outside 45 AU. This is in agreement
  with our result in the sense that we detect a compact mid-infrared dust
  emission; the dust at larger radii is cold and radiates at wavelengths
  longer than those traced by our observations. We suggest that the second
  [\ion{O}{i}] peak arises either from a dust gap or from a dust-free layer
  above the shadow of the inner rim.
 
  Following the phenomenological classification of Meeus et
  al. (\cite{meeus}), HD 101412 is a group II source, i.e. the disk is flat
  and self-shadowed in the dust. At the dust sublimation radius the disk is
  puffed-up and shadows the outer disk, which is not reached by the stellar UV
  radiation. The second peak of the [\ion{O}{i}] emission (Fig. 8) is not
  expected for such a self-shadowed geometry. We suggest that this is the
  result of a flat disk geometry for the dust and a flared geometry for the
  gas. This might be the signature of a different scale height, and hence
  vertical structure, between gas and dust beyond the inner rim. This means
  that -- in this object -- gas and dust are physically decoupled in the
  surface layers of the disk allowing the gas to emerge from the shadow of the
  inner rim. If this is confirmed, it may have important consequences for the
  disk evolution: once gas and dust decouple the dust grains settle faster
  towards the mid-plane. The dust settling favors the physical separation
  between gas and dust until all the dust settles towards the mid-plane
  (Dullemond \& Dominik \cite{dullemond04}) eventually leaving a region of
  gas in the upper layers of the disk. We thus suggest that HD 179218, HD
  135344 B and HD 101412 form an evolutionary sequence where the disk,
  initially flared, becomes flat under the combined action of gas-dust
  decoupling, grain growth and dust settling. Interestingly, Acke et
  al. (\cite{acke04}) already found that cold grains in the mid-plane of the
  disk have grown to considerably larger sizes in Meeus et al., group II
  sources than in group I sources, suggesting that disks may evolve from a
  flared to a self-shadowed geometry. Similarly, Bouwman et
  al. (\cite{bouwman}) find a correlation between the strength of the
  amorphous silicate feature and the shape of the SED, consistent with the
  settling of dust as a consequence of grain growth. What we add to this
  picture is a completely independent piece of evidence supporting this and
  perhaps the identification of the driving mechanism of this process.
  
  \begin{appendix}
    \section{Extended PAH emission}
    Emission from polycyclic aromatic hydrocarbons (PAHs) is typically detected
    towards disks around Herbig stars (e.g. Meeus et al. \cite{meeus};  Acke \&
    van den Ancker \cite{acke04}; Sloan et al. \cite{sloan} \& Geers et
    al. \cite{geers}) and in some T Tauri stars (e.g. Geers et al. \cite{geers06},
    \cite{geers}). PAHs are heated by UV-photons from the central star. In
    many cases, the PAH emission is found to be spatially extended (e.g. van Boekel
    et al. \cite{boekel}; Habart et al. \cite{habart}, \cite{habart06}; Visser et
    al. \cite{visser}; Doucet et al. \cite{doucet07} \& Boersma
    et al. \cite{boersma}). In particular, in Herbig stars, the PAH emission is
    more extended than the continuum (at similar wavelengths) and  can be as
    extended as the total size of the disk (Visser et al. \cite{visser}). In
    disks around less massive stars, given the lower UV flux, the emission is
    weaker and more compact around the central star. Three PAH emission
    features are present in the spectral range covered by our observations
    centered respectively at 8.6~$\mu$m, 11.3~$\mu$m and 12.7~$\mu$m. All the
    three sources we study here show evidence of emission of (at least one of)
    these features (e.g. van Boekel et al. \cite{boekel05} and figure 1). The
    PAH emitting region may be much larger than the continuum/silicate one (Sect
    4.2). As a consequence, at the position of the PAH emission there are (at
    least) two visibility components which contribute to the MIDI observed
    visibility. Depending on the relative contribution of PAH emission to the total
    flux at a specific wavelength, the total visibility tends to be lower due to the
    presence of the extended PAH emission. Given the contamination of telluric
    absorption corresponding to the 12.7~$\mu$m emission band, we do not
    take this feature into account in our analysis. The main findings that
    emerge from the analysis of the MIDI visibility and correlated flux are:
    
    \begin{description}
    \item{PAH 8.6~$\mu$m feature}
      
      \begin{enumerate}
      \item {\bf HD 101412}: The visibility increases monotonically between
        8~$\mu$m and 9~$\mu$m and the disk appears slightly more extended
        shortward of 9~$\mu$m than at 10~$\mu$m. This may be a signature of
        extended PAH emission\\
      \item {\bf HD 135344 B}: No evidence of 8.6~$\mu$m emission in the total
        spectrum or in any of the correlated spectra \\
      \item {\bf HD 179218}: The visibility is systematically low shortward of
        9~$\mu$m. However, it is not clear whether this just reflects the
        characteristic sinusoidal modulation of the ``ring-like'' structure or
        whether it is the PAH emission that causes this effect\\
      \end{enumerate}
      
    \item{PAH 11.3~$\mu$m feature}
      
      \begin{enumerate}
      \item {\bf HD 101412}: A drop in the visibility of HD101412 around
        11.3~$\mu$m is detected along the second baseline (59 m,
        102$\degr$). This is also the observation with the best
        signal-to-noise ratio for HD 101412. A similar trend is visible along
        the third baseline (62 m, 110$\degr$), although the large error bars
        make this detection uncertain. HD 101412 is classified as a group II
        (self-shadowed) disk. The detection of PAH emission more extended than the dust
        emitting region is peculiar for such a disk structure. The UV-excited
        PAH molecules should lie in a dust-free surface layer of the disk.
        This source has a strong [\ion{O}{i}] 6300\,\AA\ emission -- equally
        unusual for a group II source -- which led van der Plas et
        al. (\cite{plas}) to consider HD 101412 to be somewhat transitional in
        character between group I and group II. \\
      \item {\bf HD 135344 B}: A clear drop in visibility in the vicinity of
        11.3~$\mu$m is detected with MIDI with all baselines. This is even
        more clear in the correlated flux (Fig. \ref{fig:CF}): no emission
        is detected in the four measurements of the correlated flux,
        indicating that the 11.3~$\mu$m PAH emission comes from spatial scales
        larger that those resolved by MIDI.\\
      \item {\bf HD 179218}: No detectable variation of the visibility is
        visible in the MIDI observations of HD 179218 around 11.3~$\mu$m. This
        disk is classified as a group I source (flared). The flaring geometry
        and the large size found for the dust emitting region may suggest
        that in this case dust and PAHs have a similar extent.\\
      \end{enumerate}
    \end{description}
    
    \section{Inclined ring model}
    In the text, two simple geometrical models were used to fit the MIDI
    visibility: 1) a uniform disk and 2) a uniform ring. The uniform disk model is
    a first-order approximation of a spherical brightness distribution. The
    intensity ($I(\rho,\phi)$, $\rho$ and $\phi$ polar coordinates) is
    constant and non zero within a radius $\theta/2$ and null outside:
    \begin{equation}\label{I_UD}
      I(\rho,\phi) =\left\{ 
        \begin{array}{ll}
          I_0 & \qquad \rho \leq \theta/2 \\
          0 &   \qquad \rho > \theta/2
        \end{array} \right.
    \end{equation}
    given the azimuthal symmetry, $I(\rho,\phi) = I(\rho)$. The Fourier
    transform of Eq. \ref{I_UD} is:
    \begin{equation}\label{V_UD}
      V(u,v) = 2 \frac{J_1 (\pi \theta r)}{(\pi \theta r)}
    \end{equation}
    with $J_1$ Bessel function of the first order. The second model used in
    this paper is the uniform ring. In polar coordinates the brightness
    distribution of the uniform ring is:
    \begin{equation}\label{I_UR}
      I(\rho) =\left\{ \begin{array}{ll}
          0   & \qquad \rho < \theta_{in}/2\\
          I_0 & \qquad \theta_{in}/2 \leq \rho \leq \theta_{out}/2 \\
          0   & \qquad \rho > \theta_{out}/2
        \end{array} \right.
    \end{equation}
    with $\theta_{in}$ and $\theta_{out}$ inner and outer diameters of the ring.
    
    Given the linearity property of the Fourier transform, the visibility of the
    uniform ring with inner size (diameter) $\theta_{in}$ and outer size
    $\theta_{out}$ is the difference of the visibility of two uniform disks.
    The visibility of a structure inclined by an angle $i$ and rotated by an
    angle $\phi$ (position angle of the major axis) can be obtained by rotating the
    $u,v$ coordinates (e.g. Berger \& Segransan \cite{berger}):
    \begin{eqnarray}\label{eq:uv}
      u' & = & u \cdot cos(\phi) + v \cdot sin{\phi} \\
      v' & = & -u \cdot sin(\phi) + v \cdot cos{\phi}
    \end{eqnarray}
    and by applying a compression factor along the minor axis:
    \begin{equation}\label{eq:r}
      r = \sqrt{u'^2 + v'^2 \cdot cos(i)^2}.
    \end{equation}

    The visibility of an inclined uniform ring is: 
    \begin{equation}\label{eq:vring}
      V_{ring}(r) = f_{\lambda} + 2\frac{1-f_{\lambda}}{\theta_{out}^2 -
        \theta_{in}^2} \big[\theta_{out}^2\frac{J_1(\pi \theta_{out} r)}{(\pi
          \theta_{out} r)} - \theta_{in}^2 \frac{J_1(\pi \theta_{in} r)}{(\pi
          \theta_{in} r)}\big]
    \end{equation}
    where $f_{\lambda}$ is the fractional contribution of an unresolved component
    (e.g. the central star) to the total flux at wavelength $\lambda$. In the extreme 
    case of an infinitely thin ring with $\theta_{in} = \theta_{out}$, the visibility 
    becomes:
    \begin{equation}
      V_{ring}(\theta_{in}) = f_{\lambda} + (1 - f_{\lambda}) J_0(\pi \theta_{in} r).
    \end{equation}
    
    The model adopted for HD 179218 is the sum of two uniform rings:
    \begin{equation}\label{I_URUR}
      I(\rho) =\left\{ \begin{array}{ll}
          0   & \qquad \rho < \theta_1/2 \\
          I_0 & \qquad \theta_1/2 \leq \rho \leq \theta_2/2 \\
          0   & \qquad \theta_2/2 < \rho < \theta_3/2 \\
          I_0 & \qquad \theta_3/2 \leq \rho \leq \theta_4/2 \\
          0   & \qquad \rho > \theta_4/2
        \end{array} \right.
    \end{equation}
    with $\theta_1, \theta_2$, $\theta_3$ and $\theta_4$ the diameters of the two uniform rings
    respectively. The visibility of such a model is:
    \begin{equation}\label{eq:V_URUR}
      V(r) = f_{\lambda} V_{UR, in}(\theta_1, \theta_2) + (1 - f_{\lambda}) V_{UR, out}(\theta_3, \theta_4)
    \end{equation}
    with $V_{UR_in}$ and $V_{UR_out}$ the visibility of a uniform ring
    (Eq. \ref{eq:vring}) for the inner and outer component respectively.
    
  \end{appendix} 
  \begin{acknowledgements}
    D. Fedele acknowledges support by Deutscher
    Akad. Austauschdienst. (DAAD). GM acknowledges financial support by the
    Deutsche Forschungsgemeinschaft (DFG) under grant ME2061/3-1 and 3-2. We
    are grateful to the ESO staff in Garching and on Paranal for performing
    the MIDI observations in service mode. We thank the anonymous referee for
    many useful comments and suggestions which helped to improve this
    manuscript.
 \end{acknowledgements}

\end{document}